\documentclass[onecolumn,draft, 11pt]{IEEEtran}

\usepackage{color}

\usepackage{wrapfig,graphicx,tabularx,array,amsmath,amsthm,thmtools}

\usepackage{mathtools}

\usepackage{amsfonts}
\usepackage{bm}
\usepackage{bbm}
\usepackage{makecell}
\usepackage{multirow}
 \usepackage{amssymb}
\usepackage{txfonts}
\usepackage[T1]{fontenc}
\usepackage{tikz}
\usepackage[scr=dutchcal]{mathalfa}

\usepackage{cite}
\newcommand\blfootnote[1]{%
  \begingroup
  \renewcommand\thefootnote{}\footnote{#1}%
  \addtocounter{footnote}{-1}%
  \endgroup
}

\newtheorem{Theorem}{Theorem}

\newtheorem{Definition}{Definition}

\theoremstyle{definition}
\newtheorem{Example}{Example}

\begin{document}

\title{On Multiterminal Communication over MIMO Channels with One-bit ADCs at the Receivers}


\author{Abbas Khalili$^1$, Farhad Shirani$^1$, Elza Erkip$^1$, Yonina C. Eldar$^2$\\
$^1$Dept. of Electrical and Computer Engineering,
New York University, NY. \\
$^2$ Dept. of Electrical Engineering, Technion Institute of Technology, Israel\date{} }


\maketitle
\begin{abstract}
The fundamental limits of communication over multiple-input multiple-output (MIMO) networks are considered when a limited number of one-bit analog to digital converters (ADC) are used at the receiver terminals. Prior works have mainly focused on point-to-point communications, where receiver architectures consisting of a concatenation of an analog processing module, a limited number of one-bit ADCs with non-adaptive thresholds, and a digital processing module are considered. In this work, a new receiver architecture is proposed which utilizes adaptive threshold one-bit ADCs --- where the ADC thresholds at each channel-use are dependent  on  the  channel  outputs  in  the  previous  channel-uses --- to mitigate the quantization rate-loss.
Coding schemes are proposed for communication over the point-to-point and broadcast channels, and achievable rate regions are derived. In the high SNR regime, it is shown that using the proposed architectures and coding schemes leads to the largest achievable rate regions among all receiver architectures with the same number of one-bit ADCs.   \blfootnote{This work is supported by NYU WIRELESS Industrial Affiliates and
National Science Foundation grant SpecEES-1824434.}
\end{abstract}

\section{Introduction}
Next generation cellular networks  --- whether in the MHz or GHz regimes --- will utilize hundreds of antennas at the base station and in excess of ten antennas at the user terminals \cite{mo2015capacity}. A major obstacle in the implementation and adoption of  massive multiple-input multiple-output (MIMO) systems
and millimeter wave (mmWave) technologies is the high energy consumption resulting from the large number of antennas employed in the transmitter and receiver terminals \cite{rangan2014millimeter,walden1999analog}.

In traditional fully digital receivers, each antenna is connected to a distinct high resolution analog to digital converter (ADC). While ADCs are an essential component of wireless communication systems, they are a major source of power consumption \cite{rangan2014millimeter}. In conventional ADC design, the power consumption grows linearly in the number of quantization bins \cite{MIMO1}. More precisely, for an ADC with $n_q$ output bits per input symbol, the power consumption is modeled as $P=c2^{n_q}W$, where $c$ is a normalizing factor and $W$ is the sampling rate. Alternatively, if a set of $n_q$ one-bit threshold ADCs are used, then the power consumption grows linearly in $n_q$. Consequently, the use of several one-bit ADCs instead of a single high resolution ADC has been proposed to limit the power consumption at MIMO receiver terminals \cite{mo2014channel,abbasISIT2018,rini2017generalITW,koch2013low,mezghani2012capacity,mezghani2008analysis}. However, the receiver architectures studied in the literature lead to a rate-loss, in the sense that when $n_q$ one-bit ADCs are used, they achieve rates strictly less than $n_q$ bits per channel-use \cite{arxiv_PtP_ADC_2018}.

There has been a large body of work on characterizing the rate-loss due to low resolution quantization in point-to-point (PtP) MIMO systems \cite{abbasISIT2018,rini2017generalITW,koch2013low,mezghani2012capacity,mezghani2008analysis}. These works consider a PtP MIMO communication problem, where the receiver is equipped with a limited number of one-bit threshold ADCs. The receiver performs linear analog processing on the received channel output vector and feeds the resulting vector to the one-bit ADCs. The digitized output is stored and blockwise processing is performed to recover the message. In particular, \cite{abbasISIT2018} and \cite{koch2013low} show that using non-zero thresholds in the ADCs can lead to higher achievable rates. 

In a companion paper \cite{arxiv_PtP_ADC_2018}, we studied  PtP MIMO communication with one-bit ADCs at the receiver, and proposed a blockwise non-adaptive threshold receiver architecture where the ADC thresholds are fixed over the transmission block. We showed that this architecture, which uses analog delay elements, results in an achievable rate that grows linearly in the number of ADCs at high SNRs, and achieves the maximum rate for a PtP MIMO system that uses a given number of one-bit ADCs at the receiver. This is in contrast with prior works, where the increase in capacity is logarithmic in the number of one-bit ADCs \cite{abbasISIT2018,mo2015capacity}. 
However, the performance analysis in \cite{arxiv_PtP_ADC_2018} does not extend naturally to PtP communication in low SNRs and multiterminal communications. In this paper, we instead propose a class of adaptive threshold receiver architectures, where the ADC thresholds at each channel-use depend on the channel outputs in the previous channel-uses. The proposed adaptive threshold architecture resembles the
class of successive approximation register ADCs \cite{CMOS}, and is amiable to analysis for PtP communication in the low SNR regime and multiterminal communications. Furthermore, in PtP communications, the proposed adaptive threshold architecture achieves the same optimal high SNR rate as the non-adaptive threshold one of \cite{arxiv_PtP_ADC_2018}.

\begin{figure*}[t]
\centering
\includegraphics[width=1\textwidth, height=0.32\textwidth,draft=false]{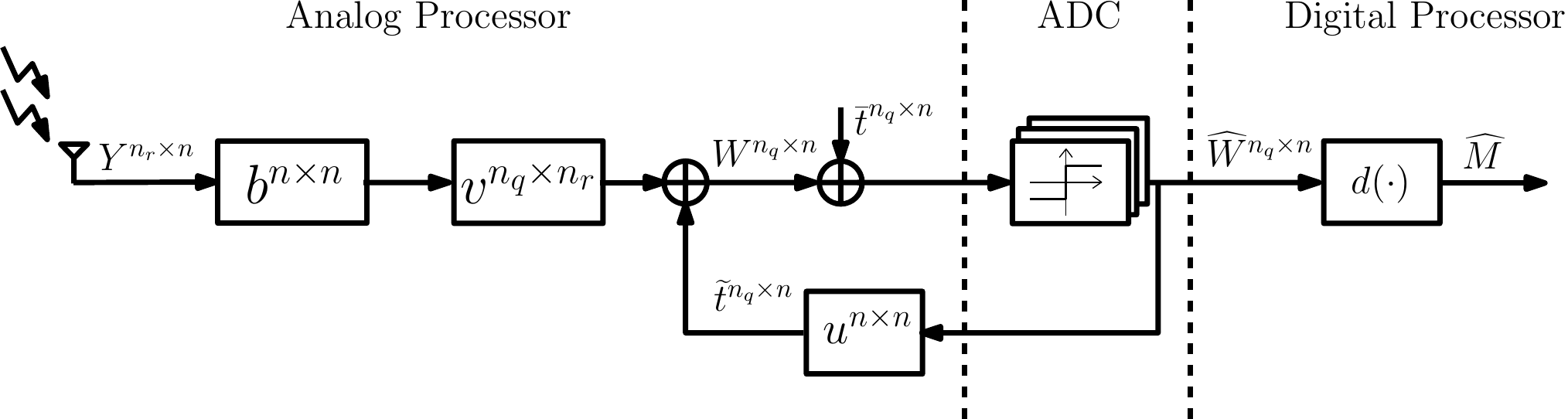}
\caption{A PtP-QMIMO system is shown where the linear combiner, temporal processing, and adaptive threshold coefficients are characterized by the matrices $v^{n_q\times n_r}$, $b^{n\times n}$, and $u^{n\times n}$, respectively. Each of the columns of $\overline{t}^{n_q\times n}$ are equal to the threshold vector $t^{n_q}$. The ADC module consists of $n_q$ one-bit ADCs.}
\label{fig:PtP}
\end{figure*}

While much of the literature focuses on PtP communication, the use of low resolution ADCs at the receivers in multiterminal communications gives rise to new challenges in interference management and successive decoding schemes. There have only been few works analyzing efficient and reliable multiterminal receiver architectures and communication strategies in the presence of low resolution ADCs. In \cite{Rassouli2018}, communication over the multiple-access channel (MAC) was studied when each transmitter is equipped with a single antenna and the receiver has a single one-bit ADC. It was shown that the optimal input distribution is discrete. Uplink communication over wireless networks modeled as a MAC, was also studied in \cite{Studer2016} and practical coding strategies were proposed when coarse quantization is used at the receiver. 

Using the adaptive threshold receiver architecture, we propose coding schemes for communication over PtP channels and the broadcast channel (BC), and
derive achievable rate or rate regions for arbitrary SNRs. We show that the proposed architecture achieves the optimal rate region in the presence of one-bit ADCs when SNR goes to infinity in all receiver terminals in the sense that the achievable region cannot be expanded by improving the architecture design without the use of additional ADCs. The ideas used in constructing the coding scheme may be used to devise coding strategies for communication over MAC as well. The MAC achievable region will appear in a longer version of this work.

The rest of the paper is organized as follows: Section \ref{Sec:Model} explains the system model. Section \ref{Sec:FBW} contains the proposed ADC module construction and receiver architectures. In Section \ref{Sec:Regions}, we derive achievable regions for various multiterminal communication settings. Section \ref{Sec:Conclusion} concludes the paper.

{\em Notation:} Sets are denoted by calligraphic letters such as $\mathcal{X}, \mathcal{U}$. The set of natural numbers and real numbers are denoted by $\mathbb{N}$ and $\mathbb{R}$, respectively. 
 The set of numbers $\{1,2,\cdots, n\}, n\in \mathbb{N}$ is represented by $[n]$. 
 For a given $n\in \mathbb{N}$, the $n$-length vector $(x_1,x_2,\hdots, x_n)$ is written as $x^n$. The subvector $(x_k,x_{k+1},\cdots,x_n)$ is denoted by $x_k^n$. We write $||x^n||_2$ to denote the $L_2$-norm of $x^n$. An $n\times m$ matrix is written as $h^{n\times m}=[h_{i,j}]_{i,j\in [n]\times [m]}$, and $I_n$ is the $n\times n$ identity matrix.
  The notation  $h^{m}(i), i\in[n]$ is used to represent the $i$th column of $h^{n \times m}$.

 
 \section{System Model}
\label{Sec:Model}
We first describe the PtP MIMO communication setup when a limited number of one-bit ADCs are available at the receiver  which also forms the basis of the BC model. Consider a PtP-MIMO system characterized by $(n_t,n_r, h^{n_r\times n_t})$, where $n_t$ is the number of transmitter antennas, $n_r$ is the number of receiver antennas, and $h^{n_r\times n_t}$ is the channel gain matrix. The input and output vector pair $(X^{n_t}, Y^{n_r})$ are related through 
\[Y^{n_r}=h^{n_r\times n_t}X^{n_t}+N^{n_r},\] where $N^{n_r}$ is a vector of independent and identically distributed  Gaussian variables with zero mean and unit variance, and the channel input has average power constraint $P$. The channel gain matrix is assumed to be fixed over the transmission block and is known at the transmitter and receiver. The receiver uses a concatenation of an analog processing module which performs analog signal processing on the received signals, an ADC module consisting of $n_q$ one-bit threshold ADCs to digitize and store the signals, and a digital processing module which performs a digital blockwise decoding operation on the stored  bits to recover the message. The communication system is called an $(n_t,n_r, h^{n_r\times n_t},n_q)$ PtP-QMIMO system. 

\textbf{One-shot Architectures:} Prior works have proposed analog processing of the received signals before quantization as a means to mitigate the rate-loss \cite{abbasISIT2018,rini2017generalITW,koch2013low,mezghani2012capacity,mezghani2008analysis}. 
In these works, communication is performed in a transmission block of length $n$. 
At each channel-use, an analog linear combiner multiplies the channel output vector $Y^{n_r}$ by a spatial analog processing matrix $v^{n_q\times n_r}$ and the output vector is fed to the threshold ADCs, where the threshold vector is denoted by $t^{n_q}$. The spatial analog processing matrix and threshold vector are assumed to be fixed over the transmission block. The ADC output vector in the $i$th channel-use is  $\widehat{W}^{n_q}(i)=Q(v^{n_q\times n_r}Y^{n_r}(i)+t^{n_q}), i\in [n]$, where $Y^{n_r}(i)$ is the channel output in that channel-use, and $Q(x^{n_q})$ is the element-wise sign quantization of $x^{n_q}$. The digital output is stored by the receiver and digital blockwise processing is performed to recover the message. Note that this architecture does not allow for analog temporal processing of the received signals prior to quantization. Consequently, it is called \textit{one-shot} receiver architecture. The one-shot capacity, maximized over all spatial analog processing matrices $v^{n_q\times n_r}$ and threshold vectors $t^{n_q}$, is denoted by $C_{OS}(h^{n_r\times n_t},n_q)$.



\textbf{Adaptive Threshold Architectures:} We propose a specific class of  adaptive threshold blockwise receiver (AT-Rx) architectures  shown in Fig. \ref{fig:PtP}. The matrix $Y^{n_r\times n}$ consists of the received channel output vectors over $n$ channel-uses, where $Y^{n_r}(i)$ is the output in the $i$th channel-use. The outputs are linearly combined by the temporal analog processing matrix $b^{n\times n}$ and the spatial analog processing matrix $v^{n_q\times n_r}$. These linear combination operations are implemented using analog linear combiners and delay elements. More precisely, in order to allow for temporal processing, the receiver uses delay elements to store the previously received signals and feeds the stored signals to the analog combiner. The analog combiner produces $v^{n_q\times n_r}Y^{n_r\times n}b^{n\times n}$.
As shown in Fig. \ref{fig:PtP}, the ADC thresholds consist of a fixed component and an adaptive component. To elaborate, let $\widehat{W}^{n_q\times n}$ be the matrix of ADC outputs, where $\widehat{W}^{n_q}(i), i\in [n]$ is the vector of ADC outputs after the $i$th channel-use.  The adaptive threshold matrix $\widetilde{t}^{n_q\times n}$ is produced by linear processing of $\widehat{W}^{n_q\times n}$ and is equal to $\widehat{W}^{n_q\times n}u^{n\times n}$, where $u^{n\times n}$ is called the adaptive threshold coefficient matrix. The vector $\widetilde{t}^{n_q}(i), i\in [n]$ is the adaptive threshold in the $i$th channel-use. The fixed component of the ADC thresholds is the vector $t^{n_q}$.  The adaptive threshold coefficient matrix $u^{n\times n}$ is strictly upper-triangular and the temporal linear combining coefficient matrix  $b^{n\times n}$ is an upper-triangular matrix. This ensures that the system can be implemented causally. 

We call the architectures described above \textit{adaptive threshold} architectures (AT-Rx). We will provide an example of an adaptive threshold architecture and justify its use in Section \ref{Sec:FBW}.
The AT-Rx architectures are formally defined below.
\begin{Definition}[{\bf AT-Rx}]
\label{Def:FBW}
Consider the PtP-QMIMO system characterized by the tuple $(n_t,n_r, h^{n_r\times n_t},n_q)$. Let $b^{n\times n}\in \mathbb{R}^{n\times n}$ be an upper-triangular matrix, and $u^{n\times n}\in \mathbb{R}^{n\times n}$ be a strictly  upper-triangular matrix. A transmission system with adaptive threshold architecture at the receiver (AT-Rx) is characterized by the tuple $(n,\Theta,v^{n_q\times n_r},{t}^{n_q}, b^{n\times n}, u^{n\times n})$, the pair of encoding and decoding functions $(e,d)$ are defined in a similar fashion as in PtP-QMIMO systems, where $\widehat{W}^{n_q}(i)=Q(v^{n_q\times n_r}\overline{Y}^{n_r}(i)+\widetilde{t}^{n_q}(i)+t^{n_q}), i\in [n]$, $\overline{Y}^{n_r\times n} = Y^{n_r\times n}b^{n\times n}$ is the temporally processed received signal, and  $\widetilde{t}^{n_q\times n}= \widehat{W}^{n_q\times n}u^{n\times n}$ is the adaptive threshold matrix. The capacity, maximized over all linear combiner matrices $v^{n_q\times n_r}$, threshold vectors $t^{n_q}$, temporal processing matrices $b^{n\times n}$ and adaptive threshold coefficient matrices $u^{n\times n}$, is denoted by $C_{AT}(h^{n_r\times n_t},n_q)$.
\end{Definition}


The BC model considered in this paper is formally defined below. An example is provided in Section \ref{Sec:Regions}.

\begin{Definition}[\bf{BC-QMIMO}]
\label{Def:BC}
 A two-user MIMO broadcast channel with one-bit ADCs (BC-QMIMO)  is characterized by the tuple $(n_t,n_{r,1},n_{r,2}, h_1^{n_{r,1}\times n_t},h_2^{n_{r,2}\times n_t},n_{q,1}, n_{q,2})$, where $n_t$ is the number of transmit antennas, $n_{r,i}$, $h_i^{n_{r,i}\times n_t}$, $n_{q,i}, i\in \{1,2\}$ are the number of receive antennas, channel gain matrix, and number of one-bit ADCs at the $i$th receiver. The channel input vector is $X^{n_t}$ and the output at the $i$th receiver is given by $Y^{n_{r,i}}= h_i^{n_{r,i}\times n_t}X^{n_t}+N^{n_{r,i}}, i\in \{1,2\}$, where $N^{n_{r,i}}$ is a vector of independent, zero-mean and unit-variance Gaussian variables.
\end{Definition}

\section{Adaptive Threshold Architectures}
\label{Sec:FBW}
It was shown in prior works that one-shot communication over PtP-MIMO systems with a limited number of one-bit ADCs at the receiver inflicts a rate-loss on the transmission system in the sense that the maximum achievable rate is strictly less than the number of one-bit ADCs even at high SNRs \cite{abbasISIT2018,rini2017generalITW,mo2015capacity}. More precisely, even in the ideal scenario when the SNR is taken to be asymptotically large,
the one-shot capacity $C_{OS}(h^{n_r\times n_t}, n_q)$ is strictly less than $n_q$. The example in Section \ref{Sec:FBW}.A  illustrates how the proposed adaptive threshold architectures eliminate the aforementioned rate-loss. Particularly, Example \ref{Ex:1}  describes a method for constructing an equivalent scalar quantizer with $2^{n_q}$ quantization bins from $n_q$ one-bit threshold ADCs. 
Section \ref{Sec:FBW}.B builds upon this motivating example to propose a general coding strategy for PtP MIMO communication with AT-Rx architectures and derive an achievable rate region for arbitrary SNRs. Section \ref{Sec:Regions} further extends these ideas to broadcast channel communications.
\subsection{Motivating Example}
The following example explains the rate-loss due to low resolution quantization in a simple SISO scenario and describes the proposed receiver architecture.

\begin{Example}
\label{Ex:1}
\textbf{One-shot Architecture:} consider a PtP SISO communication scenario (i.e. $n_t=n_r=1$), where the receiver is equipped with two one-bit ADCs (i.e. $n_q=2$). Fig. \ref{fig:simple}(a) shows a generic one-shot PtP-QMIMO receiver  in which the receiver antenna receives $Y_1$, and produces the digitized signal:
\begin{align*}
    (\widehat{W}_1,\widehat{W}_2)=
    \begin{cases}
    (-1,-1) \qquad & \text{if } Y_1<t_1,\\
    (+1,-1)&   \text{if } t_1<Y_1<t_2,\\
    (+1,+1)&   \text{if } t_2<Y_1,
    \end{cases}
\end{align*}
    where we have assumed that $t_1<t_2$ without loss of generality. Note that the symbol $(-1,+1)$ is not produced at the receiver. In the high SNR scenario, the communication noise is negligible and rates close to $\log{3}$ bits per channel-use are achievable. The rate is strictly less than $n_q=2$ bits per channel-use. 
    
\noindent    \textbf{Adaptive Threshold Architecture:} Fig.  \ref{fig:simple}(b) shows the proposed adaptive threshold architecture. The first ADC has zero threshold and outputs the sign of the received signal $Y_1$. The threshold of the second ADC is adaptive and is set to be equal to half of the output of the first ADC in the previous channel-use. In the context of Definition \ref{Def:FBW}, we have $n=2$, $v^{2\times 1}=\begin{bmatrix}
         1\\
         1
    \end{bmatrix}$, $b^{2\times 2}=I_2$, and $u^{2\times 2}= \begin{bmatrix}
    0 & -\frac{1}{2}\\ 
    0 & 0
    \end{bmatrix}.
    $ Let $(\widehat{W}_1(i), \widehat{W}_2(i))$ be the output of the two ADCs at the $i$th channel-use. Then,
    \begin{align*}
    (\widehat{W}_1(i),\widehat{W}_2(i+1))=
    \begin{cases}
    (-1,-1) \qquad & \text{if } Y_1(i)<-\frac{1}{2},\\
    (-1,+1)&   \text{if } -\frac{1}{2}<Y_1(i)<0,\\
    (+1,-1)&   \text{if } 0<Y_1(i)<\frac{1}{2},\\
    (+1,+1)&   \text{if } \frac{1}{2}<Y_1(i).
    \end{cases}
\end{align*}
 As a result, the rate of 2 bits per channel-use is achievable at high SNRs. Note that this rate is optimal in the sense that any other receiver architecture and coding scheme cannot achieve higher rates without increasing the number of one-bit ADCs.

           \begin{figure}[t]
 \centering
\includegraphics[width=0.7\textwidth, height=0.32\textwidth,draft=false]{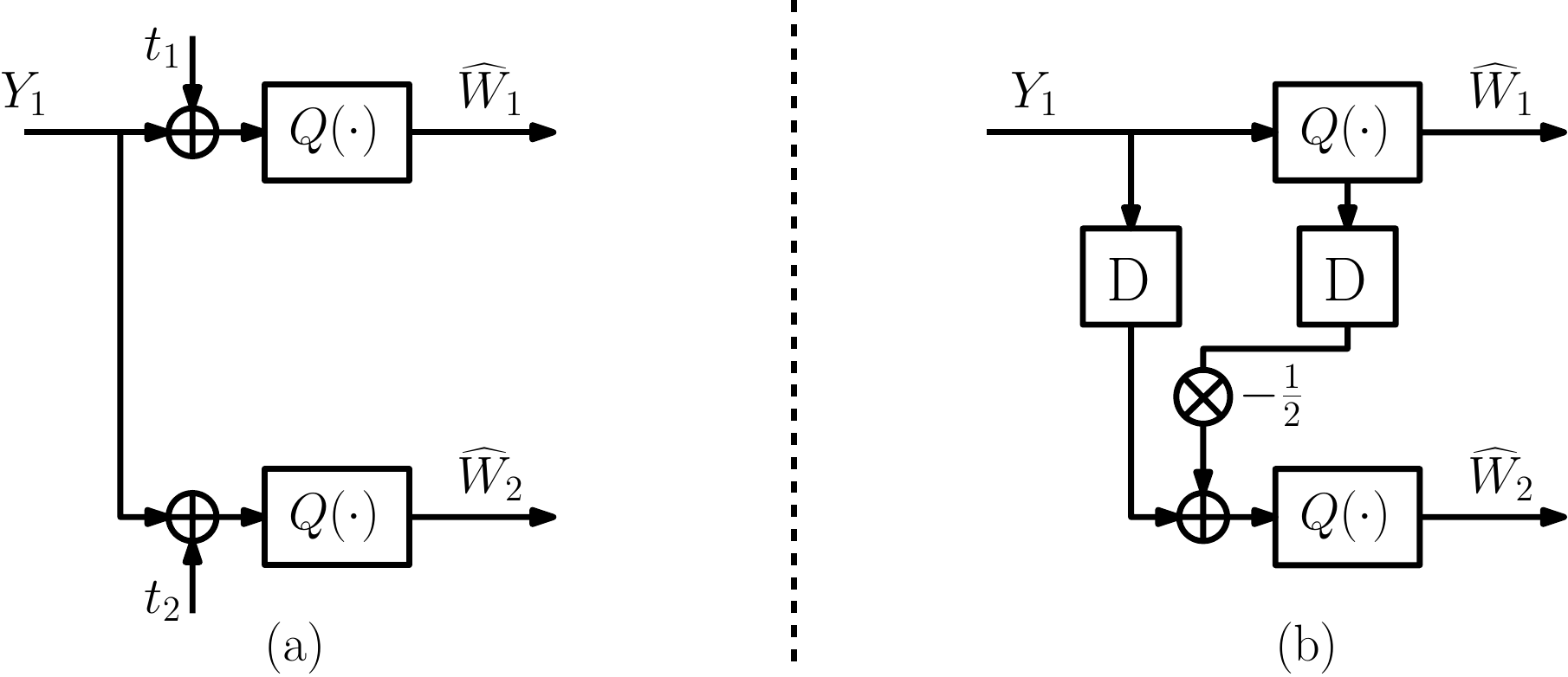}
\caption{(a) An OS-Rx system with fixed thresholds. (b) An AT-Rx system where the first  ADC has zero threshold and the second ADC has an adaptive threshold which is equal to half of the output of the first ADC.}
\label{fig:simple}
\end{figure}

\end{Example}

\subsection{Achievable PtP Rates}
In the following, we consider communication over PtP-QMIMO systems and propose a communication scheme which uses the singular value decomposition (SVD) in the analog domain to transform the communication system into $s$ parallel, non-interfering channels. The $i$th parallel channel is allocated a number of $n_{q,i}$ one-bit ADCs, where $\sum_{i\in [s]}n_{q,i}=n_q$. The adaptive threshold architecture described in Example \ref{Ex:1} is employed to construct an equivalent scalar quantizer with $2^{n_{q,i}}$ quantization bins. We calculate the maximum achievable rate for a given ADC allocation $(n_{q,1},n_{q,2},\cdots, n_{q,s})$ and power allocation $(P_1,P_2,\cdots,P_s)$ for the sub-channel subject to the total power constraint.  The following theorem describes the set of achievable rates. 

\begin{Theorem}
\label{Th:1}
For the PtP-QMIMO characterized by $(n_t,n_r, h^{n_r\times n_t},n_q)$ and average input power constraint $P$, the rate $R$ is achievable if it satisfies the following inequality:
\begin{align*}
    R \leq \max \sum_{k =1}^{s} I(\widetilde{X}_k;\widetilde{Y}_k)
\end{align*}
where the maximum is taken over $(n_{q,i})_{i\in [s]}, (P_{i})_{i\in [s]}$ such that $\sum_{i\in [s]}n_{q,i}=n_q, \sum_{i\in[s]}P_{i}=P$, $\widetilde{X}_k = a_k\cdot\left(2\widehat{X}_k-1-2^{n_{q,k}}\right), \widetilde{Y}_k = \sigma_{k} \widetilde{X}_k+N_k$, $a_k = \sqrt{\frac{3P_k}{2^{2n_{q,k}}-1}}$, $\widehat{X}_k$ is uniformly distributed over $[2^{n_{q,k}}]$, $N^s$ is a vector of i.i.d. zero-mean Gaussian variables with unit variance, and $\sigma_k$ is the $k$th singular value of $h^{n_r\times n_t}$. 


\end{Theorem}


Note that as $\textrm{SNR}\to \infty$, we have
\begin{align*}
I(\widetilde{X}_k;\widetilde{Y}_k) \stackrel{\text{(a)}}{\to} H(\widetilde{X}_k)\stackrel{\text{(b)}}{=}H(\widehat{X}_k)
\stackrel{\text{(c)}}{=} n_{q,k},    
\end{align*}
where (a) holds since at high SNR we have $h(\widetilde{X}_k|\widetilde{Y}_k)\approx 0$, (b) follows by the bijectivity of the mapping $\widetilde{X}_k = a_k\cdot\left(2\widehat{X}_k-1-2^{n_{q,k}}\right)$, and (c) holds since $\widehat{X}_k$ is uniformly distributed over $[2^{n_{q,k}}]$. As a result, the total transmission rate approaches $n_q$. Consequently, the performance converges to optimality as SNR is increased asymptotically since no more than $n_q$ bits per channel-use may be decoded using $n_q$ one-bit ADCs at each channel-use. This is in contrast with \cite{mo2015capacity} and \cite{abbasISIT2018}, where the achievable rate grows logarithmically in the number of ADCs at high SNRs. 

\section{Broadcast Channel Communication Strategies}
\label{Sec:Regions}

In this section, we propose BC coding strategies and derive achievable regions for communication over the BC. 
\begin{figure}[t]
 \centering
\includegraphics[width=0.6 \textwidth,draft=false]{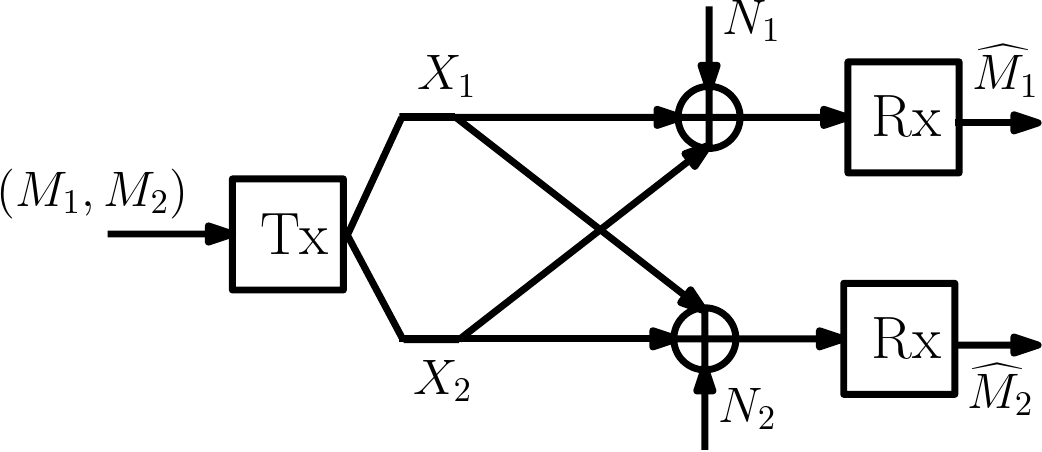}
\caption{A two user broadcast channel example, where The two users' channels are statistically equivalent. The transmitter is equipped with two antennas and each receiver is equipped with a single antenna and a one-bit threshold ADC.}
\label{fig:BC}
\end{figure}


\subsection{Motivating Example}

 The following example describes the key ideas used in the proposed coding strategies. 
\begin{Example}
\label{Ex:TS}
Consider the two-user broadcast channel shown in Fig. \ref{fig:BC}. The BC-QMIMO system is characterized by $(n_t,n_{r,1},n_{r,2}, h_1^{n_{r,1}\times n_t},h_2^{n_{r,2}\times n_t},n_{q,1}, n_{q,2})$, where $n_t=2,n_{r,1}=n_{r,2}=n_{q,1}=n_{q,2}=1$, and $h_1^{n_{r,1}\times n_t}=h_2^{n_{r,2}\times n_t}=
\begin{bmatrix}
1 & 1
\end{bmatrix}$.   We investigate the achievable rate vectors when SNR$\to\infty$ when one-shot and adaptive threshold receiver architectures are used at the receiver. 
\\\noindent\textbf{One-shot Architecture with Zero Thresholds:} if the receivers are equipped with a one-shot architectures with zero-threshold ADCs (i.e. $t_{1,k}=0, k\in \{1,2\}$), then the two users' channels are statistically equivalent. As a result, each receiver may decode the other receivers' message. Consequently, by Fano's inequality, $R_1+R_2\leq 1$.
\\\noindent\textbf{One-shot Architecture with Non-zero Thresholds:} 
consider a one-shot receiver architecture where the first receiver is equipped with a zero-threshold ADC and at the second receiver, the ADC threshold is equal to $0<\epsilon\ll P$. Let the channel input be equal to $X_1+X_2= (-1)^{U_1}\frac{\sqrt{P'}}{2}+(-1)^{U_2}\sqrt{P'}$, where $P'= \frac{4}{{5}}P$ and $U_1$ and $U_2$ correspond to the messages sent to receivers one and two, respectively. It is straightforward to verify that $R_1+R_2=1+C$, where $C$ is the capacity of a binary Z-channel with cross-over probability $\frac{1}{2}$.
\\\noindent\textbf{Adaptive Threshold Architecture:} Let each receiver be equipped with the adaptive threshold architecture, where the temporal processing matrix is $b^{2\times 2}=
\begin{bmatrix}
     1 & 1 \\
     0 & 0
\end{bmatrix}
 $, $v^{1\times 1}=1$, and the adaptive threshold coefficient matrix is $u^{2\times 2}=
\begin{bmatrix}
     0 & -\frac{1}{2} \\
     0& 0
\end{bmatrix}$. We argue that the symmetric sum-rate of $2$ bits per channel-use  ($1$ bit per channel-use for each user) is achievable using this architecture. Assume that the transmitter is to send the message $(U_{i,1},U_{i,2},\cdots, U_{i,m}), i \in \{1,2\}$ to the $i$th receiver over $m+1$ channel-uses, where $U_{i,j}, j\in  [m]$ are independent binary symmetric variables and $m$ is an odd number. The odd (even) channel-uses are used to transmit two bits of information to the first (second) receiver. More precisely, the channel input in  the $j$th channel-use is
\begin{align*}
X_j=X_{1,j}+X_{2,j}= (-1)^{U_{k,j'}}\sqrt{P'}+(-1)^{U_{k,j'+1}}\frac{\sqrt{P'}}{2},
\end{align*}
where $P'=\frac{4}{{5}}P$, in the odd channel-uses $k=1$ and $j'=j$, and in the even channel-uses $k=2$ and $j'=j-1$. We explain the decoding process at the first receiver in the first two channel-uses. The decoder receives $Y_1$ and feeds it to a zero-threshold ADC to recover $U_{1,1}$. It also stores $Y_1$ using a delay element and in the next channel-use feeds it to the ADC with threshold $\frac{1}{2}U_{1,1}$.  The second decoder uses a similar method to recover its corresponding message. Note that the resulting sum-rate approaches $2$ bits per channel-use as $m, \textrm{SNR}\to \infty$. 

\end{Example}

We observe that in the adaptive threshold architecture described in Example \ref{Ex:TS}, the transmitter uses a time-sharing method where the odd (even) channel-uses are used for transmission to the first (second). However, the one-bit ADCs at both receivers are active at all channel-uses. This leads to optimal performance at high SNRs where the communication bottleneck is the number of one-bit ADCs. In the next section, we build upon the time-sharing strategy to derive an achievable region for the general BC-QMIMO system.
\subsection{Achievable Regions}
\label{Sec:rate}
In this section, we build upon the time-sharing strategy described in Example \ref{Ex:TS} and the coding strategy used in Theorem \ref{Th:1}  provide achievable rate-regions for the two user BC-QMIMO under quantization resolution constraints when adaptive threshold architectures are used at the receivers.

\begin{Theorem}
\label{th:BC}
Consider the BC-QMIMO system characterized by $(n_t,n_{r,1},n_{r,2}, h_1^{n_{r,1}\times n_t},h_2^{n_{r,2}\times n_t},n_{q,1}, n_{q,2})$. Let $\mathcal{R}$ be the set of rate vectors $(R_1,R_2)$ for which there exists $(n_{q,{j,{k}}})_{k\in [s_j]}$ and $(P_{j,k})_{k\in [s_j]}, j\in \{1,2\}$ satisfying the following bounds:
\begin{align*}
\begin{aligned}
& \sum_{k\in [s_j]}n_{q,{j,{k}}}= n_{q,j}, \quad \sum_{k\in [s_{j}]} P_{j,k}=P,\\
   & R_j \leq \frac{1}{2}\sum_{k \in[s_j]} I(\widetilde{X}_{j,k}, \widetilde{Y}_{j,k}),
\end{aligned}
\end{align*}
where $s_j, j\in \{1,2\}$ is the number of singular values of $h_j^{n_{r,j}\times n_{t}}$, $\widetilde{X}_{j,k} = a_{j,k}\left(2\widehat{X}_{j,k}-1-2^{2n_{q,{j,k}}}\right)$, $\widetilde{Y}_{j,k} = \sigma_{j,k} \widetilde{X}_{j,k}+N_{j,k}, k\in [s_j], j\in \{1,2\}$, $a_{j,k} = \sqrt{\frac{3P_{j,k}}{2^{4n_{q,{j,k}}}-1}}$, $\widehat{X}_{j,k}$ is uniformly distributed over $2^{2n_{q,j,k}}$, and $\sigma_{j,k}$ is the $k$th singular value of $h_j^{n_r\times n_{t,j}}$. Then $conv(\mathcal{R})$ is a subset of the optimal achievable rate region, where $conv(\cdot)$ is the convex hull function.


\end{Theorem}

From Theorem \ref{th:BC}, the rate pair $(n_{q,1},n_{q,2})$ is achievable at high SNR since $I(\widetilde{X}_{j,k};\widetilde{Y}_{j,k}) \to  2n_{q,j,k}, j\in \{1,2\}, k\in [s_j]$ as shown in Section \ref{Sec:FBW}.     As a result, the coding scheme is optimal at high SNRs in the sense that the rate pair $(n_{q,1},n_{q,2})$ cannot be improved upon unless a larger number of one-bit ADCs are used at the receiver terminals.

\section{Conclusion}
\label{Sec:Conclusion}
We have studied multiterminal communication over MIMO channels when a limited number of one-bit ADCs are available at the receiver terminals. We have proposed a receiver architecture which uses adaptive thresholds ADCs to mitigate the rate-loss due to low resolution quantization. We have derived achievable rate regions for communication over point-to-point and the two user broadcast channel. We have shown that the achievable regions are tight as the SNR goes to infinity. 

\bibliographystyle{IEEEtran}
\bibliography{ref}

\appendix
\section*{Proof of Theorem \ref{Th:1}}
     Let $n_{q,1}, n_{q,2}, \cdots, n_{q,s}\in \mathbb{N}\cup \{0\}$ and $P_1,P_2,\cdots,P_s\in \mathbb{R}^{\geq 0}$ such that $\sum_{k\in [s]} n_{q,k}=n_q$ and $\sum_{k\in [s]}P_k=P$.  Let $h^{n_r\times n_t}= \Phi^{n_r\times n_r}\Sigma^{n_r\times n_t}\Gamma^{n_t\times n_t}$ be the SVD of the channel gain matrix. Let ${X}^{n_t}=\Gamma^{n_t\times n_t}\widetilde{X}^{n_t}$. The receiver receives $Y^{n_r}$ and computes $\widetilde{Y}^{n_r}$ such that $Y^{n_r}=\Phi^{n_r\times n_r}\widetilde{Y}^{n_r}$  in the analog combiner module.
     The resulting parallel channels are $\widetilde{Y}_k= \sigma_{k}\widetilde{X}_k+\widetilde{N}_k, k\in [s]$, where $\sigma_k, k\in [s]$ is the $k$th singular value of the channel matrix $h^{n_r\times n_t}$ and the power constraint $\mathbb{E}(\|\widetilde{X}^s\|_2)\leq P$ must hold.  The $k$th parallel channel is allocated $n_{q,k}$ one-bit ADCs and the average power constraint $P_k$. The $n_{q,k}$ one-bit ADCs are used to construct a uniform quantizer with $2^{n_{q,k}}$ levels as described in Example \ref{Ex:1}. More precisely, the receiver first normalizes the output of the $k$th parallel channel by dividing the output by $\frac{1}{a_k\sigma_k}$, where $a_k$ is defined  in the theorem statement. Let $\overline{Y}_{(k)}^n=(\overline{Y}_{(k),1},\overline{Y}_{(k),2},\cdots,\overline{Y}_{(k),n})$ be the normalized channel output. The relation between the input and output of the $j$th ADC at the $i$th channel-use is as follows:
     \begin{align*}
      \overline{W}_{j,i}=2^{n_{q,k}-j} Q\left(\bar{Y}_{i-j+1}-\sum_{j'=1}^{j-1}\overline{W}_{j-j',i-j'}\right), j\in [n_{q,k}],  i\in \{j,j+1,\cdots, n+j-1\}.
      \end{align*}
     where $Q(x)= 2\mathbbm{1}_{x\geq 0}-1$ is the sign function.
     The achievable rate is the summation of the rates of each sub-channel. Consider the $k$th sub-channel. It is straightforward to verify that this leads to a quantizer which is equivalent with a symmetric uniform quantizer with step size $2a_k\sigma_k$, upper and lower limit $\pm 2(2^{n_{q,k}-1}-1)a_k\sigma_k$.
     It follows from standard Shannon theoretic arguments that the transmission rate of $I(\widetilde{X}_k;\widetilde{Y}_k)$ is achievable for this channel. As a result, the total transmission rate of $\sum_{k\in [s]}I(\widetilde{X}_k;\widetilde{Y}_k)$ is achievable. 

\section*{ Proof of Theorem \ref{th:BC}}
    \label{app_sec: theorem 3 proof}
    Let $n$ be the transmission block. As described in Example \ref{Ex:TS}, the first receiver receives its corresponding messages in the odd channel-uses and the second receiver in the even channel-uses. 
     Fix  $n_{q,{j,k}}, {k\in [s_j]}, j\in \{1,2\}$ and $P_{j,k}, {k\in [s_j],j\in \{1,2\}}$ such that  $\sum_{k\in [s_j]}n_{q,j,{k}}= n_{q,j}, \quad \sum_{k\in [s_{j}]} P_{j,k}=P$, $j\in \{1,2\}$.  Let $h_j^{n_{r,j}\times n_t}= \Phi_j^{n_{r,j}\times n_{r,j}}\Sigma_j^{n_{r,j}\times n_t}\Gamma_j^{n_t\times n_t}, j\in \{1,2\}$ be the SVD of the channel gain matrices. Let ${X}_j^{n_t}=\Gamma_j^{n_t\times n_t}\widetilde{X}_j^{n_t}, j \in \{1,2\}$, where ${X}_j^{n_t}$ is the channel input at a given channel-use.    The $j$th receiver receives $Y_j^{n_{r,j}}$ and computes $\widetilde{Y}_j^{n_{r,j}}$ such that $Y_j^{n_{r,j}}=\Phi_j^{n_{r,j}\times n_{r,j}}\widetilde{Y}_j^{n_{r,j}}$ in the analog combiner module , where $j\in \{1,2\}$.
     The resulting parallel channels are $\widetilde{Y}_{j,k}= \sigma_{j,k}\widetilde{X}_{j,k}+\widetilde{N}_{j,k}, k\in [s_j], j \in \{1,2\}$, where
     $\sigma_{j,k}$ is the $k$th singular value of the channel matrix $h_j^{n_{r,j}\times n_t}$ and the power constraint
     $\mathbb{E}(\|\widetilde{X}_j^{s_j}\|_2)\leq P_j$ must hold. The $k$th parallel channel of $h_j^{n_{r,j}\times n_t}, k \in [s_j], j \in \{1,2\}$ is allocated $n_{q,j,k}$ threshold one-bit ADCs and the average power constraint $P_{j,k}$. The receiver first normalizes the output of the $k$th parallel channel by dividing the output by $\frac{1}{a_{j,k}\sigma_{j,k}}$, where $a_{j,k}$ is defined  in the theorem statement.
     We describe the quantization operation in the first receiver. The operation in the second receiver is performed in a similar manner.
     Consider the $k$th parallel channel in the first receiver. Let $\overline{Y}^{n}=(\overline{Y}_1,0,
     \overline{Y}_3,0,\overline{Y}_5,\cdots,0,\overline{Y}_{2n+1})$ be the received channel output over $n$ uses of the channel. Each $\overline{Y}_{\ell}$ is fed to one of the one-bit ADCs and the ADC operates on it for $2n_{q,1,k}$ channel-uses, such that the resulting quantizer is equivalent with a uniform scalar quantizer to $2n_{q,1,k}$ output bits. More precisely, the relation between the input and output of the one-bit ADCs at the $i$th channel-use is as follows:
     \begin{align*}
         \overline{W}_{j,i}=2^{2n_{q,1,j}-\overline{(i-2k+1)}-1}Q\left(\overline{Y}_{i-\overline{(i-2j+1)}}-\sum_{i'=i-\overline{(i-2j+1)}}^{i-1}\overline{W}_{j,i'}\right), j\in [n_{q,1,k}], i\in \{2k-1,2k,\cdots, 2n+1-\overline{n}+k+2n_{q,1,k}\},
     \end{align*}
     where $\overline{i-2j+1}$ and $\overline{n}$ are  equal to $i-2j+1$ and $n$ modulo $2n_{q,1,k}$, respectively. The achievable rate is the summation of the rates of each sub-channel. Consider the $k$th sub-channel. It is straightforward to verify that this leads to a quantizer which is equivalent with a symmetric uniform quantizer with step size $2a_{1,k}\sigma_{1,k}$, upper and lower limit $\pm 2(2^{2n_{q,1,k}-1}-1)a_{1,k}\sigma_{1,k}$.
     It follows from standard Shannon theoretic arguments that the transmission rate of $\frac{1}{2}I(\widetilde{X}_{1,k};\widetilde{Y}_{1,k})$ is achievable for this channel. As a result, the total transmission rate of $\frac{1}{2}\sum_{k\in [s]}I(\widetilde{X}_{1,k};\widetilde{Y}_{1,k})$ is achievable for the first user. The proof of achievability for the rate of the second user follows by similar arguments.
\end{document}